\documentclass[aps, preprint, groupedaddress, floatfix]{revtex4}

\usepackage{epsfig}
\usepackage{graphicx}
\usepackage{amsmath}
\usepackage{color}
\usepackage{threeparttable, tablefootnote}

\newcommand*{\SSOO}{Sr$_2$ScOsO$_6$}
\newcommand*{\SYOO}{Sr$_2$YOsO$_6$}
\newcommand*{\SYRO}{Sr$_2$YRuO$_6$}
\newcommand*{\BYRO}{Ba$_2$YRuO$_6$}

\usepackage{tabularx, booktabs}
\newcolumntype{Y}{>{\centering\arraybackslash}X}

\begin{document}

\title{Magnetically-driven orbital-selective insulator-metal transition
in double perovskite oxides}

\author{Hanghui Chen$^{1,2,3}$}

\affiliation{
 $^1$NYU-ECNU Institute of Physics, New York University Shanghai China\\
 $^2$State Key Laboratory of Precision Spectroscopy, School of Physical and Material Sciences, East China Normal University, Shanghai China\\  
 $^3$Department of Physics, New York University, New York USA\\}

\begin{abstract}
  Interaction-driven metal-insulator transitions or Mott transitions
  are widely observed in condensed-matter systems. In multi-orbital
  systems, many-body physics is richer in which an orbital-selective
  metal-insulator transition is an intriguing and unique
  phenomenon. Here we use first-principles calculations to show that a
  magnetic transition (from paramagnetic to long-range magnetically
  ordered) can simultaneously induce an orbital-selective
  insulator-metal transition in rock-salt ordered double perovskite
  oxides $A_2BB'$O$_6$ where $B$ is a non-magnetic ion (Y$^{3+}$ and 
  Sc$^{3+}$) and $B'$ a
  magnetic ion with a $d^3$ electronic configuration (Ru$^{5+}$ and
  Os$^{5+}$). The orbital selectivity originates from
  geometrical frustration of a face-centered-cubic lattice on which
  the magnetic ions $B'$ reside. Including realistic structural
  distortions and spin-orbit interaction do
  not affect the transition. The predicted orbital-selective transition 
  naturally explains the anomaly observed in the electric resistivity
  of Sr$_2$YRuO$_6$. Implications of other available experimental data
  are also discussed. Our work shows that by exploiting
  geometrical frustration on non-bipartite lattices, novel
  electronic/magnetic/orbital-coupled phase transitions can occur in
  correlated materials that are in the vicinity of metal-insulator phase
  boundary.
\end{abstract}

\maketitle

Interaction-driven metal-insulator transition (so-called Mott
transition) is one of the most striking phenomena in condensed matter
systems~\cite{Imada1998}. 
With the development of many-body methods such as dynamical 
mean field theory, we can coherently describe the Mott transition using a
single-orbital Hubbard model~\cite{Georges1996, Kotliar2006a}.

In multi-orbital systems, more complicated Mott physics emerges and
the orbital-selective Mott transition (OSMT) is a most intriguing
phenomenon~\cite{Anisimov2002}. OSMT refers to the phenomenon in which
as the transition occurs, conduction electrons become localized on
some orbitals and remain itinerant on other orbitals. The idea, which
was first introduced to explain the transport properties of
Ca$_{2-x}$Sr$_x$RuO$_4$~\cite{Anisimov2002, Nakatsuji2000, Fang2004,
  Ko2007}, has stimulated many theoretical
investigations~\cite{Koga2004, DeMedici2005, Ferrero2005, Liebsch2005,
  Biermann2005, Liebsch2007, Hoshino2017} and different mechanisms
  underlying this phenomenon have been proposed: for example different orbitals
  have different intrinisic band widths~\cite{Anisimov2002}, different on-site
energies~\cite{Werner2007}, different $p$-$d$
hybridization~\cite{Wu2008} and/or different band
degeneracies~\cite{DeMedici2009}.

In this work, we use first-principles calculations to
  introduce a new approach to induce orbital-selecitve insulator-metal
  transition in multi-orbital systems. We show that in a
multi-orbital Mott insulator with its magnetic ions residing on a
non-bipartite lattice, the occurrence of long-range magnetic ordering
can drive electrons on one orbital into a metallic state while leaving
electrons on other orbitals insulating. The orbital selectivity
originates from `geometrical frustration' of non-bipartite lattices,
which enforces some magnetic moments to be ferromagnetically coupled
in an antiferromagnetic ordering.

\begin{figure}[t]
\includegraphics[angle=0,width=0.95\textwidth]{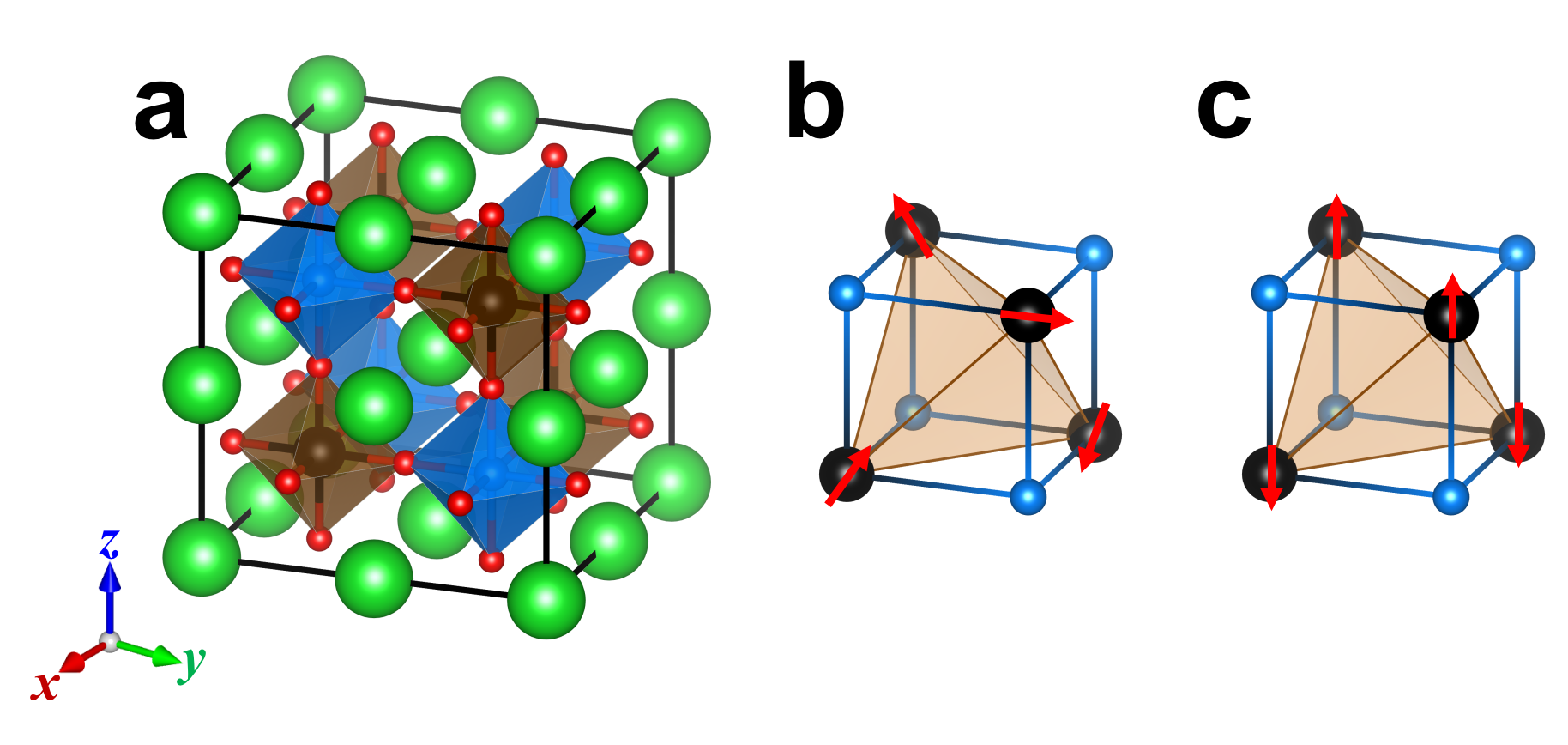}
\caption{\label{fig:atomic} \textbf{a}) A full crystal structure
  of ordered double perovskite oxides $A_2BB'$O$_6$. The blue and brown oxygen
  octahedra enclose two different types of transition metal ions $B$
  and $B'$. The green balls are $A$ ions and red balls are oxygen
  ions. \textbf{b}) and \textbf{c}) A simplified crystal structure
  of an ordered double perovskite oxide in which only transition metal
  ions $B$ and $B'$ are shown. The small blue balls are non-magnetic
  transition metal ions $B$ and the large black balls are magnetic
  transition metal ions $B'$. The red arrows denote magnetic moments of
  $B'$ ions. Panel \textbf{b}) shows a paramagnetic state in which
  magnetic moments have random orientations and fluctuate in time.
  Panel \textbf{c}) shows a type-I antiferromagnetic state in which
  magnetic moments alternate their directions between adjacent atomic
  layers along the $z$ axis
  (the ordering wave vector $\textbf{Q}=\frac{2\pi}{a} (001)$ where $a$ is the
    lattice constant).}
\end{figure}

Fig.~\ref{fig:atomic}\textbf{a} shows the crystal structure of a
rock-salt ordered double perovskite oxide $A_2BB'$O$_6$. Blue and
brown oxygen octahedra enclose two different types of transition metal
ions $B$ and $B'$. Green balls are $A$ ions and red balls are oxygen
ions. In panels \textbf{b} and \textbf{c} of Fig.~\ref{fig:atomic}, we
show a simplified structure of an ordered double perovskite oxide in
which only transition metal ions $B$ and $B'$ are shown. The small
blue balls are non-magnetic transition metal ions $B$ and the large
black balls are magnetic transition metal ions $B'$. The red arrows
denote magnetic moments of $B'$ ions. The magnetic ions $B'$ reside on
a face-centered-cubic (fcc) lattice. Panel~\textbf{b} shows a
schematic of a paramagnetic state in which magnetic moments on $B'$
ions have random orientations and fluctuate in time. Panel~\textbf{c}
shows a schematic of an antiferromagnetic state. We note that if
nearest-neighbor exchange is antiferromagnetic in nature, it is
impossible to have a `complete' antiferromagnetic ordering on a fcc
lattice in which each pair of nearest-neighbor magnetic moments is
antiferromagnetically coupled because fcc lattice has `geometrical
frustration'~\cite{Henley1987, Henley1989}.  Instead a so-called
type-I antiferromagnetic ordering is widely observed in ordered double
perovskite oxides~\cite{Aharen2009, BATTLE1989108, Carlo2013,
  Kayser2017, Taylor2015, ZAAC:ZAAC201400590, Cao2001, BATTLE1984138}. This ordering is shown
in panel \textbf{c}, in which magnetic moments alternate their
directions between adjacent atomic planes along the $z$ axis.
Mathematically the magnetic moment configuration is characterized by
an ordering wave vector $\textbf{Q}=\frac{2\pi}{a}(001)$ where $a$ is
the lattice constant. Our first-principles calculations show that
ordered double perovskite oxides which contain magnetic Ru$^{5+}$ and
Os$^{5+}$ ions are promising candidate materials which are Mott
insulators in high-temperature paramagnetic state but undergo the
aforementioned orbital-selective insulator-metal transition as the
type-I antiferromagnetic ordering occurs at low temperatures.
Experimental evidence for this transition and implications of other
available experiment data will be discussed.

The computational details of our first-principles calculations
are found in the Methods Section.

\section{Results}

\begin{table}[t]
\label{tab:material}
\begin{threeparttable}
\caption{A list of double perovskite oxides that contain Ru$^{5+}$ and
  Os$^{5+}$ in this study. AFM-I means type-I antiferromagnetic ordering.}
\begin{tabular}{cccccc}
 \hline \hline %

material  & magnetic ion & $d$ shell  & space group  & magnetic transition & ref.\\
		\hline
\BYRO & Ru$^{5+}$  &  4$d^{3}$ & ${Fm}$-${3m}$  &
                                                                       AFM-I,
                                                   $T_{N}$ $\sim$ 36
                                                   K  &
                                                        \cite{Aharen2009,
                                                        BATTLE1989108,
                                                        Carlo2013} \\
Ba$_2$ScRuO$_6$ &   Ru$^{5+}$ &4$d^3$ & ${Fm}$-${3m}$\tnote{$\dagger$} &  AFM-I, $T_{N}\sim$ 43 K
                                                                                                          &
                                                                                                            ~\cite{Kayser2017}\\
\SYRO & Ru$^{5+}$ & 4$d^{3}$ & ${P2_{1}/n}$ & AFM-I,
                                                                      $T_{\rm
                                              N}$ $\sim$ 26 K   &
                                                                \cite{Cao2001, BATTLE1984138}\\                
\SSOO & Os$^{5+}$ &  5$d^{3}$ & ${P2_{1}/n}$  & AFM-I, $T_{N}$ $\sim$
                                                92 K   &
                                                         \cite{Taylor2015}\\
\SYOO & Os$^{5+}$ &  5$d^{3}$ & ${P2_{1}/n}$  & AFM-I, $T_{N}$ $\sim$ 53 K  & \cite{ZAAC:ZAAC201400590} \\
\hline
\hline
\end{tabular}
\begin{tablenotes}\footnotesize
\item [$\dagger$] Synthesized under high pressure.
\end{tablenotes}
\end{threeparttable}
\end{table}

Table~\ref{tab:material} lists five candidate materials in this
study. In those ordered double perovskite oxides, Ru$^{5+}$ and
Os$^{5+}$ are magnetic, and Y$^{3+}$ and Sc$^{3+}$ are
non-magnetic. Both Ru$^{5+}$ and Os$^{5+}$ have a $d^3$ configuration
in which, due to Hund's rule, three $d$ electrons fill three $t_{2g}$
orbitals and form a spin $S=\frac{3}{2}$~\cite{Chen2011}. All those
four ordered double perovskite oxides exhibit type-I antiferromagnetic
ordering below N\'{e}el temperature $T_N$~\cite{Aharen2009,
  BATTLE1989108, Carlo2013, Kayser2017, Taylor2015,
  ZAAC:ZAAC201400590, Cao2001, BATTLE1984138}. For clarity, we first
study Ba$_2$YRuO$_6$ as a representative material. We discuss other
four materials in the section Discussion. Ref.~\cite{Aharen2009} shows
that Ba$_2$YRuO$_6$ crystallizes in a cubic $Fm$-$3m$ structure (space
group No. 225) and retains $Fm$-$3m$ symmetry from room temperature
down to 2.8 K (below $T_N$). The change in lattice constant due to
thermal expansion is very small ($< 0.15\%$). Experimentally, it is
found that Y$^{3+}$ and Ru$^{5+}$ site mixing is negligible or at most
very low (about $1\%$)~\cite{Aharen2009} because the size difference
between Y$^{3+}$ and Ru$^{5+}$ is significant (0.260~\AA), which
stabilizes the ordered structure~\cite{Barnes2006}. Our calculations
use its experimental low-temperature ordered structure (the details
are shown in the Supplementary Materials).

\subsection{Spectral functions}

\begin{figure}[t]
\includegraphics[angle=0,width=\textwidth]{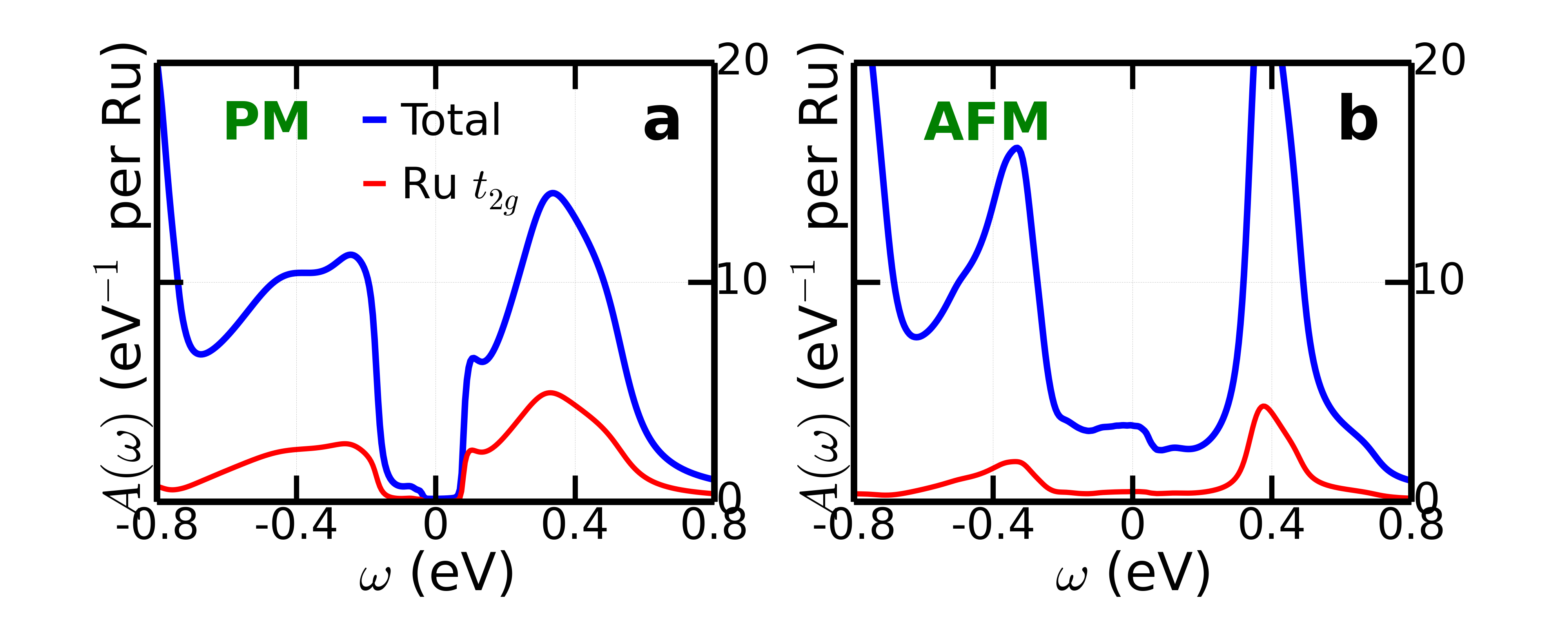}
\caption{\label{fig:dos} Spectral functions of ordered double
  perovskite Ba$_2$YRuO$_6$ calculated using DFT+DMFT with
  $U_{\textrm{Ru}} = 2.3$ eV and $J_{\textrm{Ru}} = 0.3$ eV. Panel
  \textbf{a}) shows the spectral function of paramagnetic state (PM)
  and panel \textbf{b}) shows the spectral function of type-I
  antiferromagnetic state (AFM).  The blue and red curves are total
  and Ru $t_{2g}$ projected spectral functions, respectively. The
  Fermi level is at $\omega=0$ eV. Spin up and spin down are summed
  for both PM and AFM states.}
\end{figure}

We show in Fig.~\ref{fig:dos} spectral functions of Ba$_2$YRuO$_6$ in
both paramagnetic state (panel \textbf{a}) and type-I
antiferromagnetic state (panel \textbf{b})
\cite{SingleRu}.  The blue curves are total spectral
functions and the red curves are Ru $t_{2g}$ projected spectral
functions. The paramagnetic state is insulating with a Mott gap of
about 0.2 eV. However, the type-I antiferromagnetic state shows
interesting properties: the lower and upper Hubbard bands of Ru
$t_{2g}$ states exhibit sharper peaks, compared to those in the
paramagnetic state, but the Mott gap is closed and the state is
metallic.

We first note that the transition shown in Fig.~\ref{fig:dos} is
opposite to Slater transition~\cite{Slater1951, Calder2012}. While
both transitions are driven by antiferromagnetic ordering, in Slater
transition a gap is opened in a paramagnetic metal with the occurrence
of antiferromagnetic ordering, while Fig.~\ref{fig:dos} shows that the
appearance of antiferromagnetic ordering closes the gap of a
paramagnetic insulator and induces a metallic state.

\begin{figure}[t]
\includegraphics[angle=0,width=\textwidth]{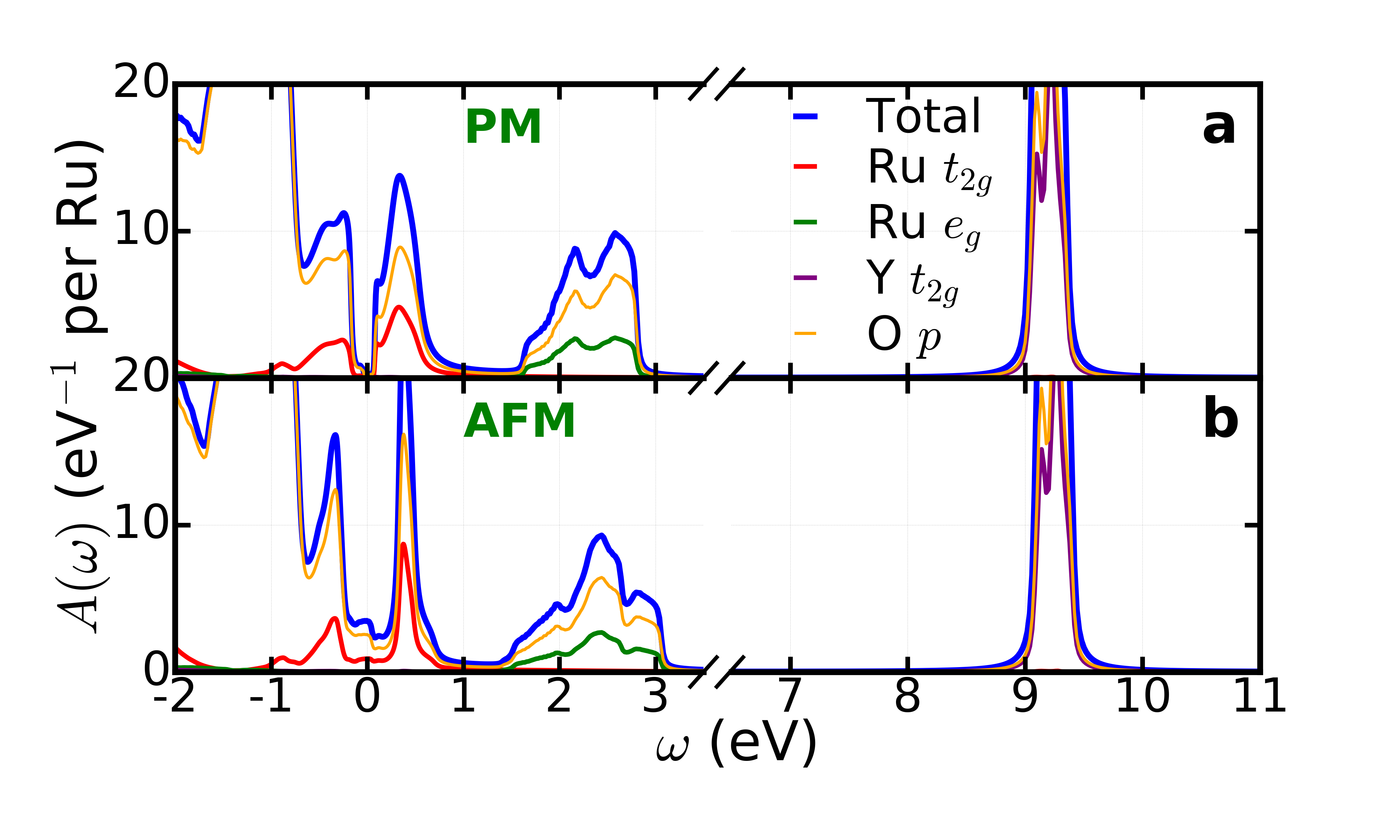}
\caption{\label{fig:dos-full} Spectral functions of ordered double
  perovskite Ba$_2$YRuO$_6$ in a large energy window (note that the
  $\omega$ axis is broken). Panel \textbf{a}) shows the spectral
  function of paramagnetic state (PM) and panel \textbf{b}) shows the
  spectral function of type-I antiferromagnetic state (AFM).  The blue
  curves are total spectral functions; the red curves are Ru $t_{2g}$
  projected spectral functions; the green curves are Ru $e_{g}$
  projected spectral functions; the purple curves are Y $t_{2g}$
  states; the orange curves are O $p$ projected spectral
  functions. The Fermi level is at $\omega=0$ eV. Spin up and spin
  down are summed for both PM and AFM states.}
\end{figure}

Second, we show that the gap closing has nothing to do with charge
transfer between Ru$^{5+}$ and Y$^{3+}$ ions~\cite{Chen2013a,
  Kleibeuker2014a}. In Fig.~\ref{fig:dos-full}, we show the spectral
functions of Ba$_2$YRuO$_6$ in a larger energy window. In addition to
total and Ru $t_{2g}$ projected spectral functions, we also show Ru
$e_g$ projected spectral function (green), Y $t_{2g}$ projected
spectral function (purple) and O $p$ projected spectral function
(orange). We find that Ru $e_g$ states have higher energy than Ru
$t_{2g}$ states due to crystal field splitting, and Y $t_{2g}$ state
have even higher energy than Ru $e_g$ states. This is consistent with
the nominally empty $d$ configuration of Y$^{3+}$.  We note that even
in plain DFT-PBE calculations (without Hubbard $U$), Y $t_{2g}$ states
have higher energy than Ru $t_{2g}$ and $e_g$ states (see Fig. 1 in
the Supplementary Materials). This indicates that there is \textit{no}
charge transfer between Y$^{3+}$ and Ru$^{5+}$ ions in both
paramagnetic and type-I antiferromagnetic states of Ba$_2$YRuO$_6$.

\subsection{Orbital-selective transition}

\begin{figure}[t]
\includegraphics[angle=0,width=\textwidth]{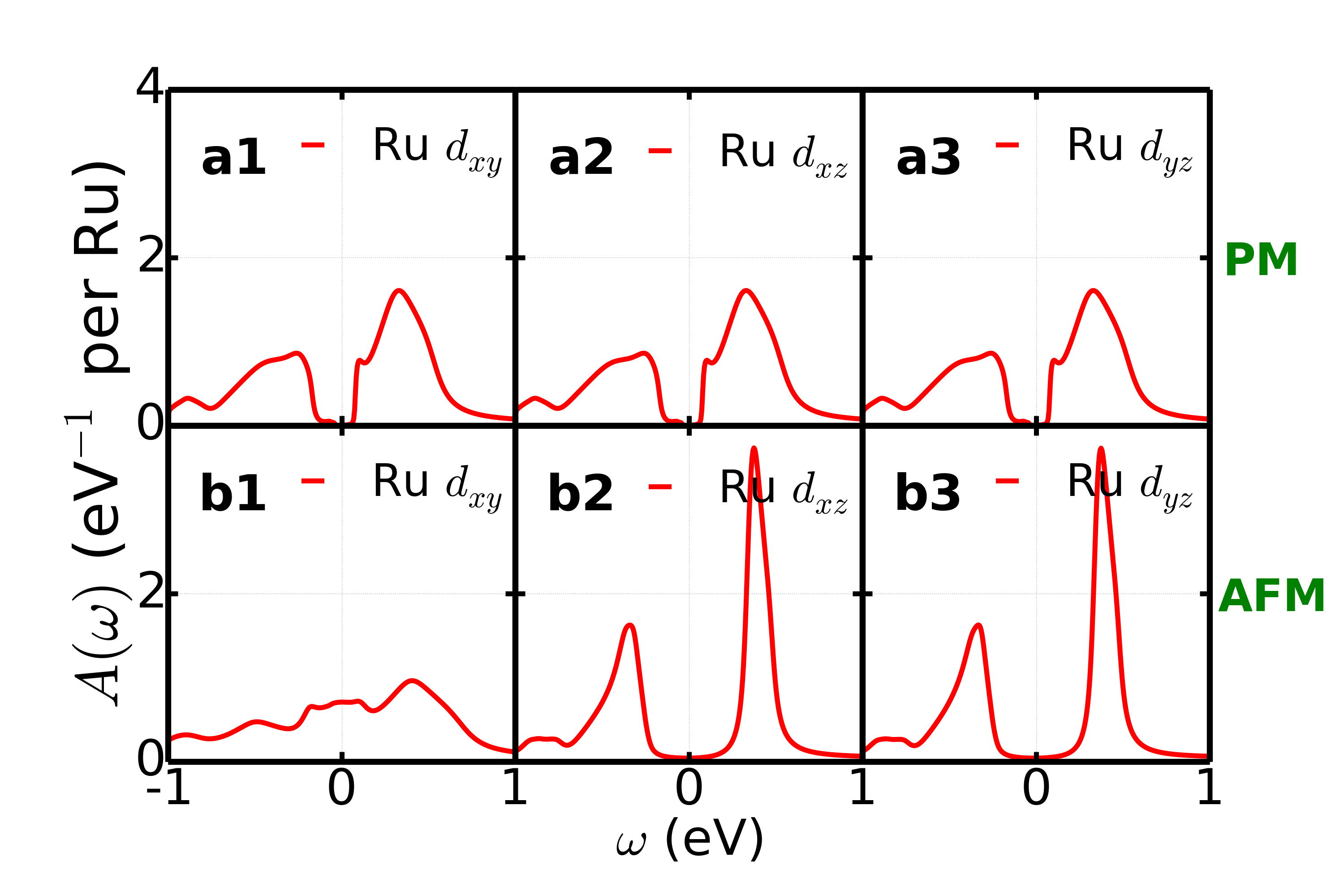}
\caption{\label{fig:dos-orbital} Spectral functions of double
  perovskite Ba$_2$YRuO$_6$ in paramagnetic state (PM,
  panels \textbf{a}) and in type-I antiferromagnetic state (AFM,
  panels \textbf{b}). The ordering wave vector is
  $\textbf{Q}=\frac{2\pi}{a} (001)$ where $a$ is the lattice constant.
  Panels \textbf{a1}) and \textbf{b1}) Ru $d_{xy}$ projected spectral
  functions. \textbf{a2}) and \textbf{b2}) Ru $d_{xz}$ projected spectral
  functions. \textbf{a3}) and \textbf{b3}) Ru $d_{yz}$ projected spectral
  functions. The Fermi level is at $\omega=0$ eV. Spin up and spin down
  are summed for both PM and type-I AFM states.}
\end{figure}

In this section, we show that the gap closing in Ba$_2$YRuO$_6$ is
driven by the orbital-selective insulator-metal transition as we
mentioned in the Introduction. Fig.~\ref{fig:dos-orbital} is the key
result, in which we decompose the spectral function of Ba$_2$YRuO$_6$
into three Ru $t_{2g}$ orbital projections, in the paramagnetic state
and in the type-I antiferromagnetic state (the ordering wave vector
$\textbf{Q}=\frac{2\pi}{a}(001)$).  We can see that in the
paramagnetic state, three Ru $t_{2g}$ orbitals have identical
projected spectral functions due to cubic symmetry. A small Mott gap
is opened up in the paramagnetic state. However, in the type-I
antiferromagnetic state, three Ru $t_{2g}$ orbitals have different
projected spectral functions. Ru $d_{xy}$ orbital exhibits metallic
property with the gap closed, in contrast to Ru $d_{xy}$ orbital in
the paramagnetic state (column \textbf{1} of
Fig.~\ref{fig:dos-orbital}). On the other hand, Ru $d_{xz}$ and Ru
$d_{yz}$ orbitals show stronger insulating property with the gap size
increased and the peaks of lower/upper Hubbard bands becoming sharper
(columns \textbf{2} and \textbf{3} of Fig.~\ref{fig:dos-orbital}).

\begin{figure}[t]
\includegraphics[angle=0,width=\textwidth]{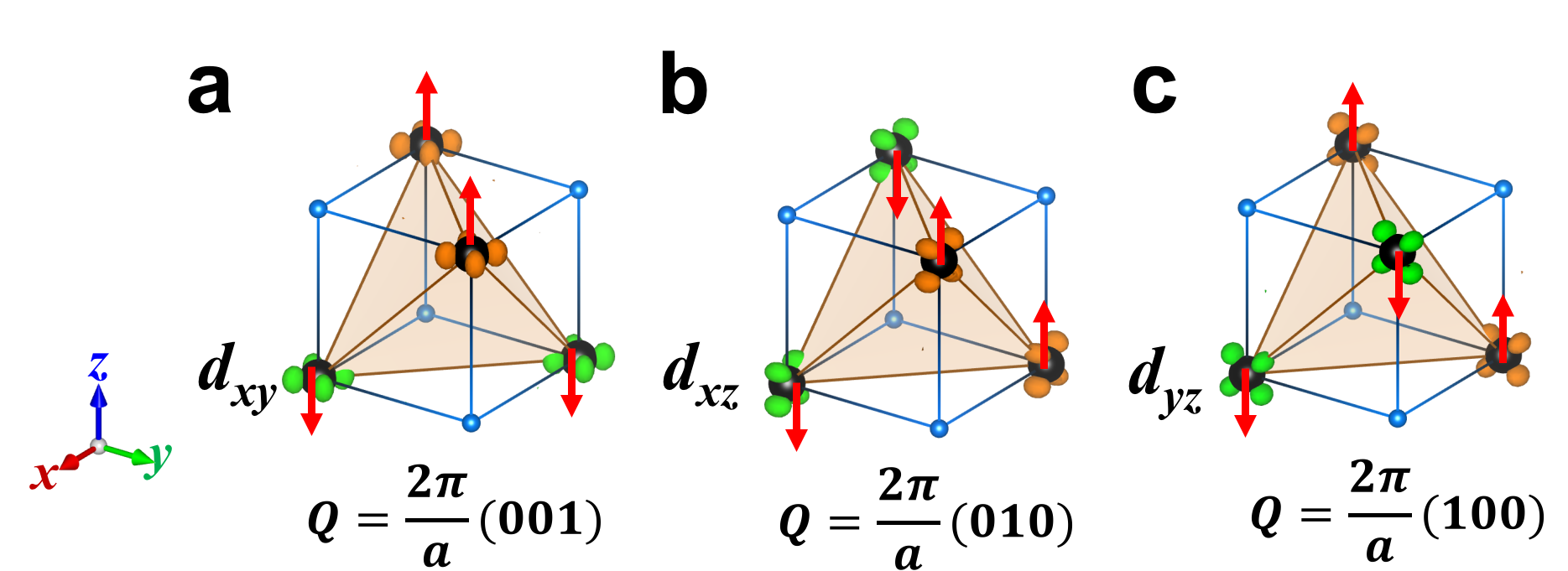}
\caption{\label{fig:orbital} Different magnetic configurations in
  ordered double perovskite Ba$_2$YRuO$_6$. The red arrows denote Ru
  $t_{2g}$ magnetic moments. For clarity, only Ru (large black balls)
  and Y (small blue balls) are explicitly shown. The orange and green
  shades are spin-resolved iso-value surface of integrated local
  spectral function for Ba$_2$YRuO$_6$ in the type-I antiferromagnetic
  state. The integral runs from $E_F - 0.05$ eV to $E_F$ where $E_F$
  is the Fermi level. The orange (green) color indicates the spectral
  function of spin up (down). Panel \textbf{a}): the ordering wave
  vector $\textbf{Q}=\frac{2\pi}{a}(001)$ and the integrated local
  spectral function close to the Femi level shows a $d_{xy}$
  character; panel \textbf{b}): the ordering wave vector
  $\textbf{Q}=\frac{2\pi}{a}(010)$ and the integrated local spectral
  function close to the Femi level shows a $d_{xz}$ character; panel
  \textbf{c}): the ordering wave vector
  $\textbf{Q}=\frac{2\pi}{a}(100)$ and the integrated local spectral
  function close to the Femi level shows a $d_{yz}$ character.}
\end{figure}

The orbital-selectivity, i.e. which Ru $t_{2g}$ orbital undergoes the
insulator-metal transition with the occurrence of type-I
antiferromagnetic ordering is related to the Ru magnetic moment
configuration, which is characterized by the ordering wave vector
$\textbf{Q}$.  For type-I antiferromagnetic ordering, there are three
ordering wave vectors: $\textbf{Q}=\frac{2\pi}{a} (001)$,
$\frac{2\pi}{a}(010)$ or $\frac{2\pi}{a}(100)$ where $a$ is the
lattice constant. They correspond to different axes along which Ru
magnetic moments alternate their directions between adjacent atomic
planes. As is shown in Fig.~\ref{fig:orbital}, for each ordering wave
vector $\textbf{Q}$, Ru magnetic moments are parallel in $\frac{1}{3}$
of nearest-neighbor Ru pairs and are anti-parallel in the other
$\frac{2}{3}$ of nearest-neighbor Ru pairs. The Ru magnetic moments
that are parallel single out a plane and the Ru $t_{2g}$ orbital that
lies in the plane (rather than out of the plane) undergoes an
insulator-metal transition. For example, in
Fig.~\ref{fig:orbital}\textbf{a}, the ordering wave vector
$\textbf{Q}=\frac{2\pi}{a} (001)$ and the parallel Ru magnetic moments
single out $xy$ plane. Together we show an iso-value surface, which is
the spin-resolved (orange and green) integrated local spectral
function around the Fermi level~\cite{Supple}.  The shape of the
iso-value surface clearly indicates that the many-body density of
states close to the Fermi level has a $d_{xy}$ character, which is
consistent with Fig.~\ref{fig:dos-orbital}. In
Fig.~\ref{fig:orbital}\textbf{b} and \textbf{c}, we repeat the
calculations with different ordering wave vectors
$\textbf{Q}=\frac{2\pi}{a}(010)$ and $\frac{2\pi}{a}(100)$.  As we
change $\textbf{Q}$, the states at the Fermi surface show $d_{xz}$ and
$d_{yz}$ orbital character, respectively. This partial `ferromagnetic
coupling' in the type-I antiferromagnetic ordering is the key to
explain the orbital-selective insulator-metal transition. In
Fig.~\ref{fig:orbital}\textbf{a}, the Ru magnetic moments are
ferromagnetically coupled in the $xy$ plane and antiferromagnetically
coupled in the $xz$ and $yz$ planes.  The largest hopping matrix
element for Ru $d_{xy}$ orbital is the one in the $xy$ plane between
the Ru nearest-neighbors. In the $xy$ plane, the parallel Ru magnetic
moments facilitate scattering upon excitation and thus increase
coherence and band width for Ru $d_{xy}$ orbital~\cite{Salamon2001,
  example}.  If the band width is large enough, the Mott gap is closed
for the Ru $d_{xy}$ orbital, which is exactly what
Fig.~\ref{fig:dos-orbital}\textbf{b1} shows.  Similarly, for Ru
$d_{xz}$ ($d_{yz}$) orbital, the largest hopping matrix element is the
one in $xz$ ($yz$) plane, but in that plane the Ru $t_{2g}$ magnetic
moments are anti-parallel, which hinders scattering upon excitation
and thus decreases band width and further increases band
gap~\cite{Salamon2001, example}. We note in Fig.~\ref{fig:dos-orbital}
that compared to the paramagnetic state, in the type-I
antiferromagnetic state, the gaps of Ru $d_{xz}$ and $d_{yz}$ orbitals
are indeed larger and the peaks of lower/upper Hubbard bands of Ru
$d_{xz}$ and $d_{yz}$ orbitals become sharper. Applying the same
analysis to different magnetic configurations in
Fig.~\ref{fig:orbital}\textbf{b} and \textbf{c} shows that Ru $d_{xz}$
($d_{yz}$) undergoes the insulator-metal transition with the
occurrence of type-I antiferromagnetic ordering of
$\textbf{Q}=\frac{2\pi}{a}(010)$ ($\textbf{Q}=\frac{2\pi}{a}(100)$).
We emphasize here that because both paramagnetic state and
antiferromagnetic state in Fig.~\ref{fig:dos-orbital} are calculated
at the same low temperature, it indicates that the occurrence of
type-I antiferromagnetic ordering is the driving force to induce the
orbital-selective insulator-metal transition.

\subsection{Electric conductivity}

\begin{figure}[t]
\includegraphics[angle=0,width=0.7\textwidth]{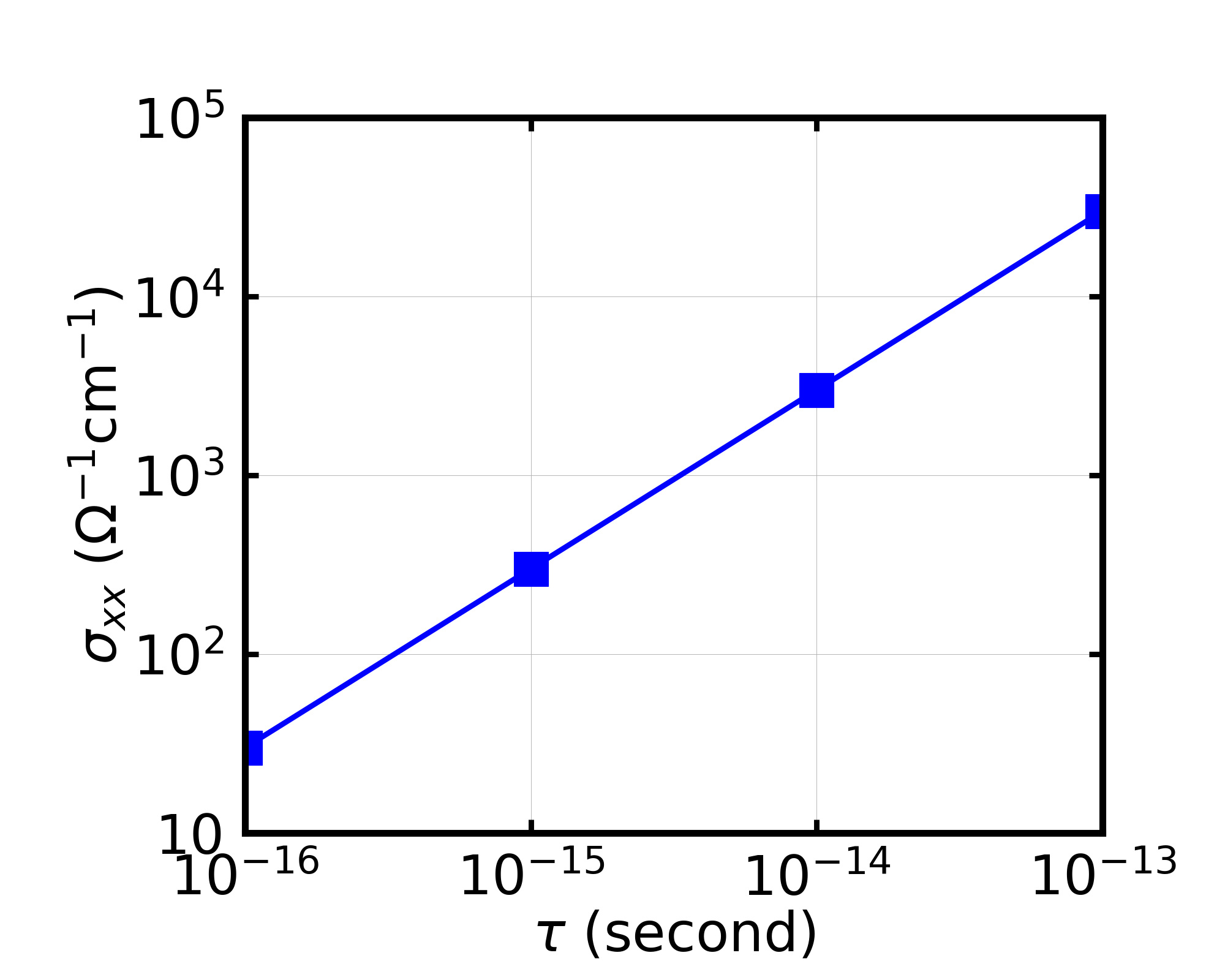}
\caption{\label{fig:rho} Electric conductivity
    ($\sigma_{xx}$ component) of Ba$_2$YRuO$_6$ in the type-I
    antiferromagnetic state (given the ordering wave vector
    $\frac{2\pi}{a}(001)$) as a function of relaxation time $\tau$,
    calculated by DFT+$U$ method ($U_{\textrm{Ru}} = 2.3$ eV and
    $J_{\textrm{Ru}} = 0.3$ eV). For simple metals, $\tau$ is around
    $10^{-14}$ s.}
\end{figure}
  
A direct consequence of the electronic structure shown in
Fig.~\ref{fig:dos-orbital} is anisotropic transport properties of
Ba$_2$YRuO$_6$ in a type-I antiferromagnetic state. We calculate
electric conductivity using DFT+$U$ method within
linear response theory and semi-classical approximation
framework~\cite{Linear, thesis}. We explain that for long-range magnetically
ordered states, because the self-energy is small and its frequency
dependence is weak, DFT+DMFT and DFT+$U$ methods yield very similar
results.

In DFT+$U$ method, electric conductivity origins from intra-band
transitions, which can be calculated from band structure.
Using linear response theory and semi-classical approximation, we have
~\cite{Linear, thesis}:

\begin{equation}
\sigma_{\alpha\beta} = \frac{4\pi e^2}{V}\sum_{n\textbf{k}}2\left(-\frac{\partial f(\epsilon)}{\partial \epsilon}\right)_{\epsilon=\epsilon_{n\textbf{k}}}\left(\textbf{e}_{\alpha}\cdot\frac{\partial \epsilon_{n\textbf{k}}}{\partial \textbf{k}}\right) \left(\textbf{e}_{\beta}\cdot\frac{\partial \epsilon_{n\textbf{k}}}{\partial \textbf{k}}\right)\tau
\end{equation} 
where $f(\epsilon)$ is the Fermi-Dirac distribution, $\alpha, \beta = x,y,z$
and $\tau$ is the
relaxation time. Note that $\tau$ is not directly calculated by
DFT+$U$ method, but is treated as a parameter. Our calculations find
that the off-diagonal components of electric conductivity
vanish due to crystal symmetry
(Ba$_2$YRuO$_6$ has a $Fm$-$3m$ structure). The diagonal components
of electric conductivity have two independent values:
$\sigma_{xx}=\sigma_{yy}$ and
$\sigma_{zz}$. This is because type-I antiferromagnetic ordering
breaks cubic symmetry (given $\frac{2\pi}{a}$(001) ordering wave
vector). Anisotropy in electric conductivity arises from the fact that
in a type-I antiferromagnetic state (given $\frac{2\pi}{a}$(001)
ordering wave vector), Ru $d_{xy}$ orbital is metallic while Ru
$d_{xz}$ and $d_{yz}$ orbitals are insulating. This means that
intra-band transitions contribute to $\sigma_{xx}$ and $\sigma_{yy}$,
but not to $\sigma_{zz}$. Our calculations find a finite
electric conductivity $\sigma_{xx} = \sigma_{yy}$ (see Fig.~\ref{fig:rho})
and a vanishing electric conductivity $\sigma_{zz} = 0$.

\subsection{Magnetic energetics}

We have shown that an orbital-selective insulator-metal transition can
occur in ordered double perovskite Ba$_2$YRuO$_6$ as the material
transitions from the paramagnetic state into the type-I
antiferromagnetic (AFM) state with decreasing temperatures. While
type-I AFM ordering has been observed in experiment (see
Table~\ref{tab:material}), as a self-consistent check, we calculate
other types of long-range magnetic orderings: ferromagnetic ordering
(FM) and antiferromagnetic ordering with magnetic moments alternating
directions along (111) axis (the ordering wave vector
$\textbf{Q}=\frac{2\pi}{a}\left(\frac{1}{2}\frac{1}{2}\frac{1}{2}\right)$
and we refer to it as type-II AFM) ~\cite{Supple}. We use DFT+$U$
method (with the same $U_{\textrm{Ru}}$ and $J_{\textrm{Ru}}$) to
calculate the energy difference between these magnetic orderings
because technically i) DFT+$U$ method can calculate larger
  systems than DFT+DMFT method (we need an 80-atom cell to calculate
  type-II antiferromagnetic ordering~\cite{Vasala2015}); ii) DFT+$U$
  method can achieve much higher accuracy than CTQMC-based DFT+DMFT
  method~\cite{Gull2011}.  Due to the quantum Monte Carlo nature of
  CTQMC algorithm, the accuracy we can obtain from DFT+DMFT method is
  on the order of 10 meV per cell.  DFT+$U$ method can converge a
  total energy of 1 meV per cell accuracy or even higher.  In
  addition, as we have mentioned in the previous section, DFT+DMFT and
  DFT+$U$ methods produce consistent results for long-range ordered
  states. That is the physical reason why we may alternatively use
  DFT+$U$ method to calculate the total energy for magnetically
  ordered states.

Using type-I AFM state as the reference, we find FM and type-II AFM
are higher in energy than type-I AFM by 110 meV/f.u. and 37 meV/f.u.,
respectively. The result that FM has higher energy than type-I
AFM shows that the nearest-neighbor exchange coupling is indeed
antiferromagnetic in nature. The reason that type-I AFM is more stable
than type-II AFM is because in type-I AFM state, for each Ru magnetic
moment, $\frac{2}{3}$ of its nearest-neighbor magnetic moments are
anti-parallel and the other $\frac{1}{3}$ of its nearest-neighbor
magnetic moments are parallel; in type-II AFM state, for each Ru
magnetic moment, half of its nearest-neighbor magnetic moments are
anti-parallel and the other half are parallel. Since the nearest-neighbor Ru
exchange coupling is intrinsically antiferromagnetic, and type-I AFM
ordering has more antiferromagnetic coupled nearest-neighbor pairs of
Ru magnetic moments than type-II AFM ordering, this explains why
type-I AFM ordering is more stable.  Our results are consistent with
the experimental measurements~\cite{BATTLE1989108, Aharen2009, Carlo2013}.

We note that the fcc lattice on which the magnetic ion Ru resides has
`geometrical frustration', therefore complicated magnetic orderings
(non-collinear and/or non-coplanar etc.) are possible in the ground
state~\cite{TIWARI2013, Chen2010b, Chen2011}. However, at finite
temperatures, by the mechanism of `order by disorder', collinear
magnetic orderings are favored by thermal
fluctuations~\cite{Henley1987, Henley1989} and collinear type-I AFM
ordering is indeed observed in experiments~\cite{BATTLE1989108,
  Aharen2009, Carlo2013}.  In our current study, it is the
\textit{first} long-range magnetic ordering which emerges from a
paramagnetic state that is relevant to the orbital-selective
insulator-metal transition.

\subsection{Spin-orbit interaction}

\begin{figure}[t]
\includegraphics[angle=0,width=\textwidth]{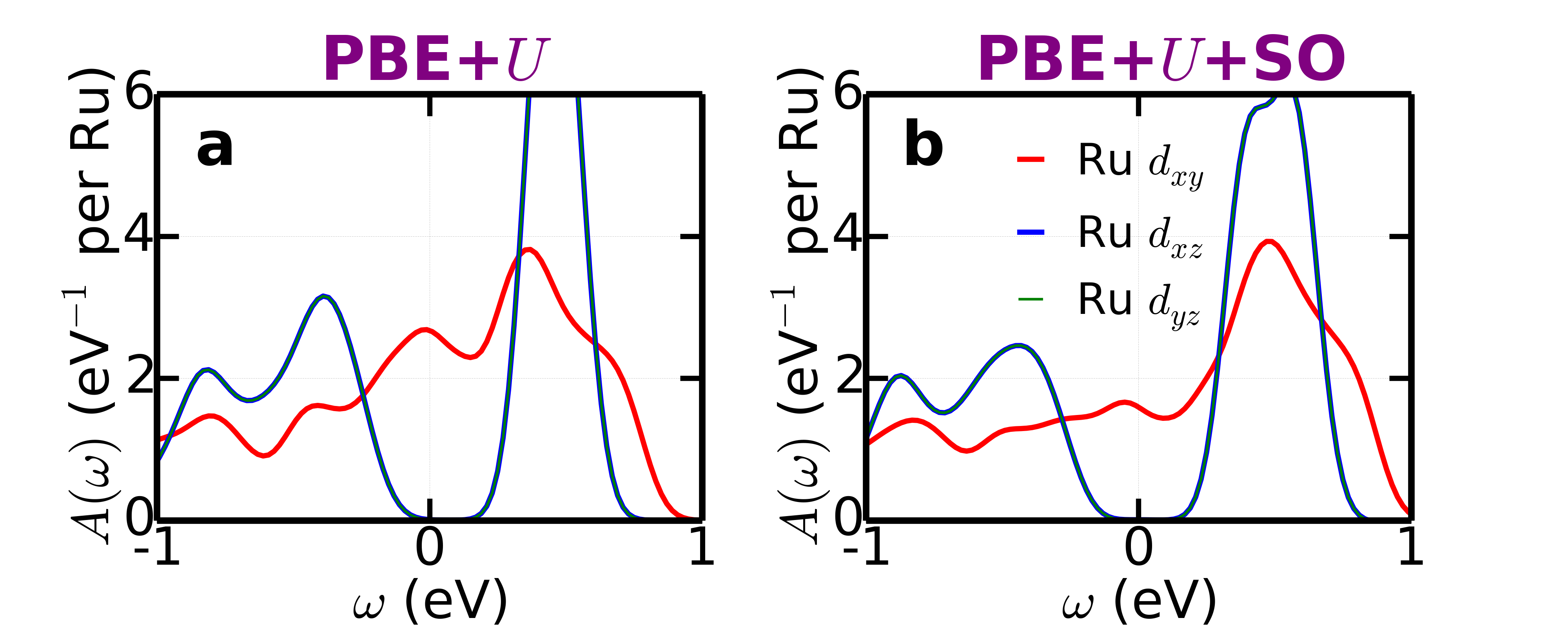}
\caption{\label{fig:spin-orbit} Spectral function $A(\omega)$ for type-I
  antiferromagnetic Ba$_2$YRuO$_6$ with the ordering wave vector
  $\textbf{Q}=\frac{2\pi}{a}(001)$ where $a$ is the lattice constant. 
  The Fermi level is at $\omega=0$ eV.
  Panel \textbf{a}) calculated by DFT+$U$ method and panel \textbf{b}) is calculated by
  DFT+$U$+SO method. The red, blue and green lines are Ru $d_{xy}$, Ru $d_{xz}$
  and Ru $d_{yz}$ projected spectral functions, respectively. For DFT+$U$ calculations,
  spin up and spin down are summed up. For DFT+$U$+SO calculations, the
  spins are aligned along the $z$ axis.}
\end{figure}

We notice that Ru has $4d$ orbitals and spin-orbit (SO) interaction
plays a more pronounced role in $4d$ and $5d$ magnetic ions than $3d$
magnetic ions. In this section, we discuss whether spin-orbit
interaction may affect the magnetically-driven orbital-selective
insulator-metal transition.

We note that currently DFT+DMFT+SO method is not feasible in
multi-orbital systems because
spin-orbit interaction induces an intrinsic sign problem in the CTQMC
algorithm~\cite{Gull2011}. But we find that in the antiferromagnetic
(AFM) ordered state, the frequency dependence in the self energy is
much weaker than that in the paramagnetic state~\cite{Supple}. This
indicates that Hartree-Fock approximation is as good as DMFT to
describe the AFM ordered state. Therefore we compare DFT+$U$ and
DFT+$U$+SO methods. 

In the presence of spin-orbit interaction, spin is
directly coupled to crystal lattice. In type-I AFM state with
$\textbf{Q}=\frac{2\pi}{a}(001)$, we globally rotate all Ru magnetic moments
in real space and find that they are stabilized along the $z$ axis. 

Fig.~\ref{fig:spin-orbit} shows the spectral functions for type-I
AFM state of Ba$_2$YRuO$_6$ (with an ordering wave vector
$\textbf{Q}=\frac{2\pi}{a}(001)$), calculated using DFT+$U$ method
(panel \textbf{a}) and DFT+$U$+SO method (panel \textbf{b}). The red,
blue and green lines are the spectral functions projected
onto Ru $d_{xy}$, Ru $d_{xz}$ and Ru $d_{yz}$ orbitals,
respectively. Using both methods, we find that with the
ordering wave vector $\textbf{Q}=\frac{2\pi}{a}(001)$, Ru $d_{xz}$ and
Ru $d_{yz}$ orbitals are insulating while Ru $d_{xy}$ orbital is
metallic. This orbital-dependent feature is also consistent with the
spectral function calculated by DFT+DMFT method
(Fig.~\ref{fig:dos-orbital}).

This result is in fact not surprising because in the current study,
the magnetic ions of double perovskite oxides have a $d^3$
configuration. Due to Hund's rule, the three electrons fill three
$t_{2g}$ orbitals and form a spin $S=\frac{3}{2}$ state. The
orbital degree of freedom is completely quenched and the system is
presumably well described by a spin-only Hamiltonian~\cite{Chen2010b,
  Chen2011}. Therefore including spin-orbit interaction does not
significantly change the electronic structure, as is shown in
Fig.~\ref{fig:spin-orbit}.

\subsection{Phase diagram with Hubbard $U$}

In the previous sections, we use a single value of Hubbard
$U_{\textrm{Ru}}$ to perform all the calculations. Now we
  discuss the phase diagram as a function of Hubbard $U_{\textrm{Ru}}$
  with $J_{\textrm{Ru}}$ fixed at 0.3 eV,
  calculated by DFT+DMFT method. We find that there are two critical
values of Hubbard $U$ (see Fig.~\ref{fig:criticalU}): i) as
$U>U_{c1}$, both the high-temperature paramagnetic state (PM) and the
low-temperature type-I antiferromagnetic state (AFM) are insulating;
ii) as $U<U_{c2}$, both the high-temperature PM and low-temperature
AFM states are metallic and iii) as $U_{c2} < U < U_{c1}$, the
high-temperature PM state is insulating and the low-temperature AFM
state is metallic. It is precisely in the region of $U_{c2} < U <
U_{c1}$ that the magnetically-driven orbital-selective insulator-metal
transition can occur at the magnetic critical temperature $T_N$.  For
Ba$_2$YRuO$_6$, we find $U_{c1} = 3.2$ eV and $U_{c2} = 1.5$ eV. While
the accurate value of Hubbard $U$ for Ru is yet to be determined, the
range set by $U_{c1}$ and $U_{c2}$ is achievable for a $4d$ transition
metal ion. We also
  note that Fig.~\ref{fig:criticalU} shows two types of phase
  transition. One is the AFM-metallic to
  AFM-insulating transition on the Hubbard $U$ axis (at low temperatures).
  The other is the
    PM-insulator to AFM-metal state transition as temperature decreases.
  Both types of
  phase transition are continuous. The
  $U$-driven phase transition is continuous because increasing $U$
  gradually separates the majority and minority spins of Ru $d_{xy}$
  orbital (given a $\frac{2\pi}{a}$(001) ordering wave vector) and
  eventually opens a gap. The PM-insulator to AFM-metal
  transition is continuous
  too, because the gap closing of Ru $d_{xy}$ orbital (given a
  $\frac{2\pi}{a}$(001) ordering wave vector) is achieved by gradually
  aligning the Ru $d_{xy}$ moments and increasing the band width of Ru
  $d_{xy}$ orbital till the majority and minority spins of Ru $d_{xy}$
  orbital overlap in energy.


\section{Discussion}

\begin{figure}[t]
\includegraphics[angle=0,width=0.8\textwidth]{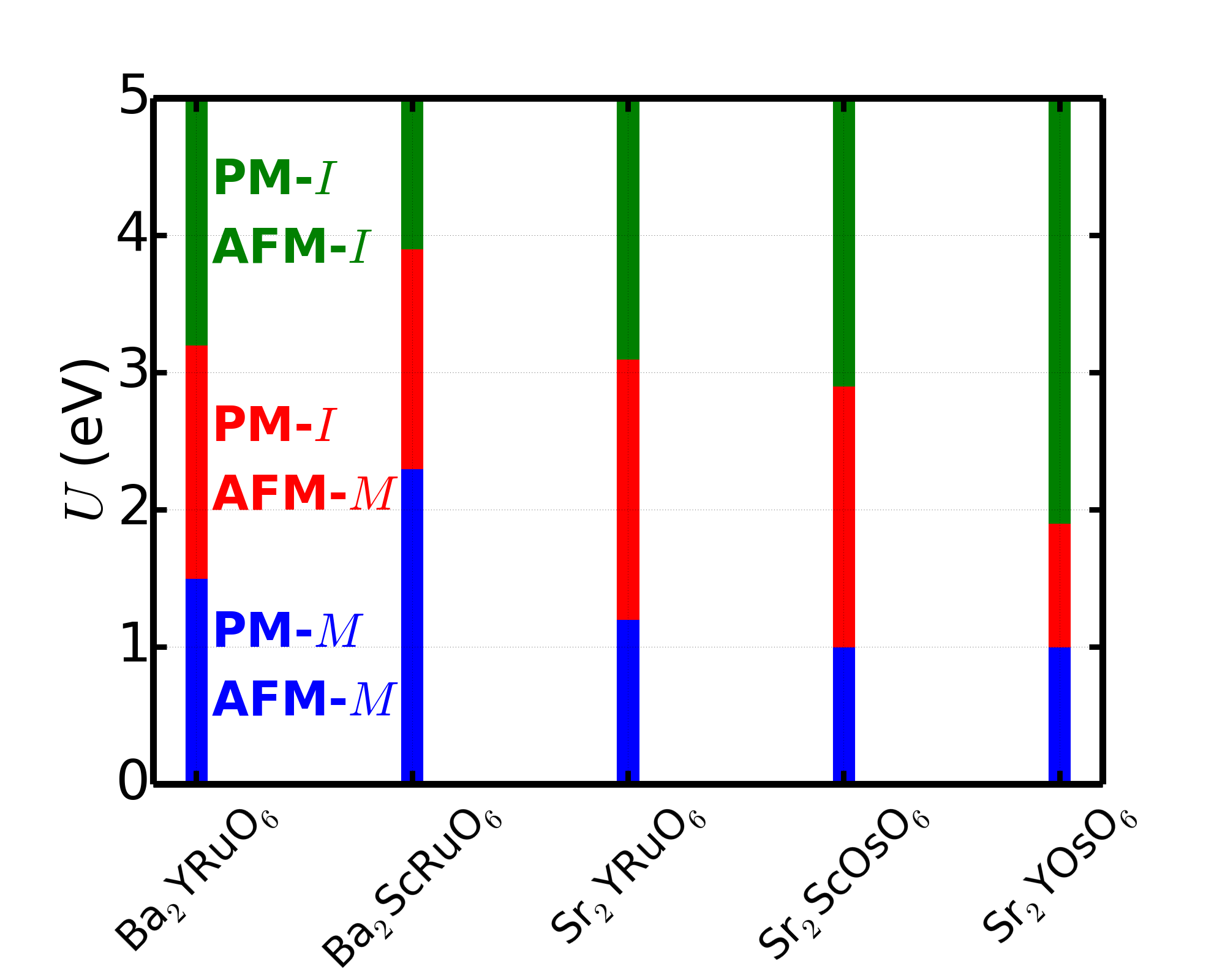}
\caption{\label{fig:criticalU} Phase diagram of Ba$_2$YRuO$_6$,
  Ba$_2$ScRuO$_6$, Sr$_2$YRuO$_6$, Sr$_2$ScOsO$_6$ and Sr$_2$YOsO$_6$
  as a function of Hubbard $U$ on magnetic ions Ru$^{5+}$ and
  Os$^{5+}$. The Hund's $J_{\textrm{Ru}}$ and $J_{\textrm{Os}}$ are
  fixed at 0.3 eV. There are two critical values of $U$. As
  $U>U_{c1}$ (green part), both high-temperature paramagnetic (PM) and
  low-temperature type-I antiferromagnetic (AFM) states are
  insulating. As $U<U_{c2}$ (blue part), both high-temperature PM and
  low-temperature type-I AFM states are metallic. As $U_{c2} < U <
  U_{c1}$, high-temperature PM state is insulating and low-temperature
  type-I AFM state is metallic. The critical values of $U_{c1}$ and
  $U_{c2}$ (in unit of eV) are (3.2, 1.5) for Ba$_2$YRuO$_6$, (3.9, 2.3)
  for Ba$_2$ScRuO$_6$, (3.1, 1.2) for Sr$_2$YRuO$_6$, (2.9, 1.0)
  for Sr$_2$ScOsO$_6$ and (1.9, 1.0) for Sr$_2$YOsO$_6$.}
\end{figure}

We have provided a comprehensive study on the magnetically-driven
orbital-selective insulator-metal transition in Ba$_2$YRuO$_6$ in
the section Results. However, the transition is not unique to
Ba$_2$YRuO$_6$; it is general to ordered double perovskite oxides with
one $d^3$ magnetic ion and one non-magnetic ion as long as the
material is a Mott insulator that lies close to the metal-insulator
phase boundary in the paramagnetic state.

In this section we study other four ordered double perovskite oxides
that are listed in Table~\ref{tab:material} and discuss the connection 
of our theoretical results to the available experimental data.
Ba$_2$ScRuO$_6$, Sr$_2$YRuO$_6$, Sr$_2$ScOsO$_6$ and 
Sr$_2$YOsO$_6$ have been
synthesized and their experimental structures (used in the
calculations) are shown in the Supplementary Materials. 
  We use DFT+DMFT method to calculate the $U$ phase diagram for these
  four double perovskite oxides (the Hund's $J_{\textrm{Ru}}$ and
  $J_{\textrm{Os}}$ are fixed at 0.3 eV~\cite{Dang2015, Han2016,
  Vecchio2016}). The results are shown in Fig.~\ref{fig:criticalU}.
Like Ba$_2$YRuO$_6$, Ba$_2$ScRuO$_6$ also crystallizes in the cubic
$Fm$-$3m$ structure (space group No. 225 $Fm$-$3m$)~\cite{Kayser2017}. However,
the lattice constant of Ba$_2$ScRuO$_6$ is smaller than that of
Ba$_2$YRuO$_6$ by about 2\%~\cite{Kayser2017, Aharen2009}, which leads
to larger hopping matrix elements. Therefore the critical Hubbard
$U_{c1}$ and $U_{c2}$ for Ba$_2$ScRuO$_6$ are both larger than those
for Ba$_2$YRuO$_6$.  On the other hand, Sr$_2$YRuO$_6$, 
Sr$_2$ScOsO$_6$ and Sr$_2$YOsO$_6$ all
crystallize in a distorted structure (space group No. 14
$P2_1/n$)~\cite{Cao2001, BATTLE1984138, Taylor2015,
  ZAAC:ZAAC201400590}.  Due to rotations and tilts of RuO$_6$ and
OsO$_6$ oxygen octahedra, metal-oxygen-metal bond angle is
smaller than that in a cubic
structure~\cite{angle}. This results in reduced hopping and therefore
the critical Hubbard $U_{c1}$ and $U_{c2}$ for all three double
perovskite oxides are smaller than those for Ba$_2$YRuO$_6$.
We note that while in our calculations there is uncertainty about the
accurate value of Hubbard $U$ on transition metal ions (Ru$^{5+}$ and
Os$^{5+}$), different `iso-electronic' materials (see
Table~\ref{tab:material}) provide a fairly large range of $U$ in which
the predicted transition can occur (shown in Fig.~\ref{fig:criticalU}).

Next we turn to available experimental data. Magnetic properties of
the five materials listed in Table~\ref{tab:material} have been
carefully studied~\cite{BATTLE1989108, Carlo2013, Kayser2017,
  Taylor2015, ZAAC:ZAAC201400590, Cao2001, BATTLE1984138}. Type-I
antiferromagnetic ordering has been observed in all these double
perovskite oxides. Remarkably, Cao \textit{et. al.} observes a sharp
anomaly in the electric resistivity $\rho(T)$ of Sr$_2$YRuO$_6$ at the
magnetic ordering temperature
$T_N$~\cite{Cao2001}. Ref.~\cite{Cao2001} measures two types of
resistivity: $\rho_{ab}(T)$ for the basal plane and $\rho_c(T)$ for
the out-of-plane $c$-axis. As the temperature $T$ is above the
N\'{e}el temperature $T_N$, both $\rho_{ab}(T)$ and $\rho_c(T)$
exhibit insulating properties: they rapidly increase as the
temperature decreases. However, just below $T_N$, $\rho_{ab}(T)$
exhibits a clear anomaly: it changes the sign of its slope and slowly
decreases with lowering temperatures (a metallic-like
behavior). Interestingly, this anomaly is only evident in
$\rho_{ab}(T)$ but is absent in $\rho_c(T)$. $\rho_c(T)$ exhibits
insulating property both above and below $T_N$ with a weak ``kink''
feature at $T_N$. Just below $T_N$, $\rho_c(T)$ increases slightly
faster with decreasing temperatures than it does just above $T_N$.
Our predicted phase transition provides an explanation
  for the anomaly observed in the resistivity of Sr$_2$YRuO$_6$ at
  $T_N$. Considering that the magnetic ordering wave vector is along
the $c$-axis ~\cite{Cao2001}, the anomaly in $\rho_{ab}(T)$ shows that
Ru $d_{xy}$ orbital (which lies in the $ab$ plane) undergoes an
insulator-metal transition at $T_N$ (see panel $\textbf{1}$ of
Fig.~\ref{fig:dos-orbital}). On the other hand, Ru $d_{xz}$ and
$d_{yz}$ orbitals remain insulating at $T_N$ and therefore the anomaly
is not observed in $\rho_c(T)$. The gap size associated with Ru
$d_{xz}$ and $d_{yz}$ orbitals increases at $T_N$ (see panels
$\textbf{2}$ and $\textbf{3}$ of Fig.~\ref{fig:dos-orbital}), which
explains the ``kink'' behavior at $T_N$.

However, as the temperature further decreases,
  $\rho_{ab}$ of Sr$_2$YRuO$_6$ undergoes a second phase transition
  from an antiferromagnetic metal to an antiferromagnetic
  insulator~\cite{Cao2001}.  According to the authors of
  Ref.~\cite{Cao2001}, the second phase transition arises from the
  fact that Dzyaloshinskii-Moriya interaction (DM-interaction) cants
  Ru spins and induces weak ferromagnetism, which eventually reopens
  the gap.

The second phase transition is interesting by itself and deserves 
further investigation, but is outside the scope of our current study. 
In our calculations, we consider type-I antiferromagnetic state (no
weak ferromagnetism) in all material candidates.

Because Sr$_2$YRuO$_6$ has a distorted structure and the
presence of DM-interaction complicates the analysis of transport
measurements, we suggest that a very similar compound Ba$_2$YRuO$_6$
is a cleaner system to observe our predicted phase transition.
Ba$_2$YRuO$_6$ has a cubic structure (space group $Fm$-$3m$)
and inversion symmetry of $Fm$-$3m$ space group forbids
DM-interaction. Without the second phase transition, $\rho_{ab}$
should show a turning-point at $T_N$ (this has already been observed in
Sr$_2$YRuO$_6$) and then monotonically decrease with decreasing
temperatures.

Another cleaner material candidate is probably Ba$_2$ScRuO$_6$, which
also crystallizes in a $Fm$-$3m$ structure. Ref.~\cite{Kayser2017} shows that
in double perovskite Ba$_2$ScRuO$_6$, a double-kink feature is observed
in its magnetic susceptibility, which indicates two ordering
temperatures ($T_N = 31$ and 44 K). However, only one peak is observed
in its heat capacity, which corresponds to the higher ordering
temperature.  The origin of the transition at the lower ordering
temperature is not clear. A measurement of low-temperature electric
resistivity for Ba$_2$ScRuO$_6$ is desirable, which will probe the
predicted orbital-selective transition and may help unlock the puzzle
of two ordering temperatures.

Finally, we mention that in
order to observe the transition, we need the material to lie close to
the metal-insulator phase boundary in the paramagnetic state (but
still on the insulating side). Therefore, for $3d$ transition metal
ions such as Mn$^{4+}$ ($d$ shell configuration $3d^3$), because a
typical $U$ is about 4 to 5 eV (larger than all the $U_{c1}$
calculated), we do not expect to observe the orbital-selective
insulator-metal transition in $3d$ double perovskite oxides such as
Sr$_2$TiMnO$_6$. For $4d$, $5d$ transition metal ions such as
Ru$^{5+}$ and Os$^{5+}$, because the Hubbard $U$ gets smaller and the
metal $d$ band width gets larger, complex oxides that contain $4d$ and
$5d$ transition metal ions are much closer to metal-insulator phase
boundary in paramagnetic state and therefore they are more promising
candidate materials to observe the transition we predict here.

\section{Conclusion}

In conclusion, we use first-principles calculations to demonstrate a
magnetically-driven orbital-selective insulator-metal transition in
ordered double perovskite oxides $A_2BB'$O$_6$ with a non-magnetic ion
$B$ (Y$^{3+}$ and Sc$^{3+}$) and a $d^3$ magnetic ion $B'$ (Ru$^{5+}$
and Os$^{5+}$). With decreasing temperatures, as the material
transitions from paramagnetic insulating (Mott) state to type-I
antiferromagnetic (AFM) state, one $t_{2g}$ orbital of the magnetic
ion becomes metallic while the other two $t_{2g}$ orbitals of the
magnetic ion become more insulating. The origin of the transition
arises from `geometric frustration' of a fcc lattice, which enforces
some magnetic moments to be ferromagnetically coupled in an
antiferromagnetic ordering. The orbital-selectivity is associated with
the ordering wave vector $\textbf{Q}$ of type-I AFM state.  We hope
our study can stimulate further experiments to provide more compelling
evidence for the predicted electronic phase transition in ordered
double perovskite oxides that contain $4d$ and $5d$ transition metal
ions.

\section{Methods}

We perform first-principles calculations by using density functional
theory (DFT)~\cite{Hohenberg1964, Kohn1965} plus Hubbard $U$
correction (DFT+$U$)~\cite{Liechtenstein1995}, DFT plus Hubbard $U$
correction and spin-orbit interaction (DFT+$U$+SO)~\cite{Takeda1978}
and DFT plus dynamical mean field theory
(DFT+DMFT)~\cite{Kotliar2006}.  Both DFT+$U$ and DFT+$U$+SO methods
are implemented in the Vienna Ab-initio Simulation Package
(VASP)~\cite{Kresse1996, Kresse1996a}. In DMFT method, a
continuous-time quantum Monte Carlo algorithm (CTQMC)~\cite{Werner2006} is
used to solve the impurity problem~\cite{Gull2011}.  The impurity
solver was developed by K. Haule's group at Rutgers
University~\cite{Haule2007}. In DMFT calculations, both
paramagnetic and antiferromagnetic states for all material candidates
are computed at a temperature of 116 K. Convergence of key results is
checked at 58 K and no significant changes are observed in electronic
structure.

For long-range magnetically ordered calculations using DFT+$U$,
DFT+$U$+SO and DFT+DMFT as well as paramagnetic calculations using
DFT+DMFT, a non-spin-polarized exchange correlation functional is used
in the DFT component~\cite{Park2015, Chen2016a}.  The spin symmetry is
broken by the Hubbard $U$ and Hund's $J$ interactions.

Electronic structures are calculated using DFT+DMFT method. Magnetic
energy differences are calculated using DFT+$U$ method and effects of
spin-orbit (SO) coupling are studied by using DFT+$U$+SO method.

In the DFT part, we use generalized gradient approximation with
Perdew-Burke-Ernzerhof (PBE) parametrization~\cite{Perdew1996} for the
exchange correlation functional. For DFT+DMFT calculations, the
correlated metal $d$ orbitals and the oxygen $p$ orbitals are
constructed using maximally localized Wannier
functions~\cite{Marzari2012}. As for the interaction
  strengths, we first use one set of interaction parameters
  $U_{\textrm{Ru}}$ = 2.3 eV and $J_{\textrm{Ru}}$ = 0.3 eV to show
  the representative electronic structure and then study Hubbard $U$
  dependence. We show that the transition we predict can occur in a
  range of interaction strength for all candidate materials. We note
  that recent calculations of SrRu$_2$O$_6$~\cite{Streltsov2015, Hariki2017,
    Okamoto2017} show that for a $t_{2g}$-$p$ model,
  $U_{\textrm{Ru}}$ is about 5 eV from constrained random-phase-approximation
  (cRPA)~\cite{Okamoto2017, Hariki2017}, which is larger than the upper
  limit $U_{c1}$ below which our predicted transition can be
  observed. However, the ``kink'' observed in the resistivity of
  Sr$_2$YRuO$_6$ indicates that the antiferromagnetic ordered state of
  Sr$_2$YRuO$_6$ exhibits metal-like behavior around $T_N$, implying
  that the interaction strength $U_{\textrm{Ru}}$ in Sr$_2$YRuO$_6$
  might be smaller than that in SrRu$_2$O$_6$ probably due to
  different crystal structure, or single-site DMFT method with a cRPA
  value of interaction strength may favor the insulating phase. This
  deserves further study in future work.

In DFT+$U$, DFT+$U$+SOC and DFT+DMFT calculations, we use
  a charge-only exchange correlation functional (i.e. not depending on
  spin density) in the DFT component. A charge-only double counting is
  also used in all methods.  Previous works show that this choice can
  avoid an unphysically large exchange-splitting in spin-dependent
  exchange correlation functionals~\cite{ChenJia2015, Park2015,
    Chen2016a}.

More computational details are found in the Supplementary Materials.

\section{Data Availability Statement}
The data that support the findings of this study are available from
the corresponding author upon reasonable request.

\section{Competing financial interests}
The authors declare no competing financial interests or non-financial interests.

\section{Author contribution}

H. Chen conceived the project, performed first-principles calculations,
analyzed data and wrote the manuscript.

\begin{acknowledgments}
  We are grateful to useful discussion with Andrew J. Millis, Jernej
  Mravlje, Sohrab Ismail-Beigi, Gang Chen, Yuan Li and
  Hongjun Xiang. The work is funded by National Science Foundation of
  China under the grant No. 11774236. Computational facilities are
  provided via Extreme Science and Engineering Discovery Environment
  (XSEDE) resources and National Energy Research Scientific Computing
  Center (NERSC).
\end{acknowledgments}

\appendix

\clearpage
\newpage


\begin{thebibliography}{65}
\bibitem[{exa()}]{example}
\bibinfo{note}{{A simple material with a half-filled $t_{2g}$ shell is studied
  in the Supplemenatray Materials to demonstrate the correlation between
  long-range magnetic ordering and band widths.}}
\bibitem[{ang()}]{angle}
\bibinfo{note}{{In Sr$_2$YRuO$_6$, Sr$_2$ScOsO$_6$ and Sr$_2$YOsO$_6$, the
  average metal-oxygen-metal bond angle is about 160$^{\circ}$, while in
  Ba$_2$YRuO$_6$ and Ba$_2$ScRuO$_6$, the metal-oxygen-metal bond angle is
  180$^{\circ}$.}}
\end{thebibliography}

\end{document}